\documentclass[aps,showpacs,preprint]{revtex4}
\usepackage{graphicx}
\begin{document}
\title{Multiple Round Quantum Dense Coding And Its Implementation Using
Nuclear Magnetic Resonance \footnote{Corresponding authors: Jingfu
Zhang, zhang-jf@mail.tsinghua.edu.cn  and G. L. Long,
gllong@mail.tsinghua.edu.cn}}
\author{\small{} Jingfu Zhang$^{1,2}$, Jingyi Xie$^{3}$,  Chuan Wang$^{1,2}$,
Zhiwei Deng$^{4}$, Zhiheng Lu,$^{5}$and Guilu Long$^{1,2,6}$}

\address{$^{1}$Key Laboratory For Quantum Information and Measurements of MOE,  and
Department of Physics, Tsinghua University, Beijing,
100084, P R China\\
$^{2}$Center For Quantum Information, Tsinghua University, Beijing
100084, P R China\\
$^{3}$Department of Materials Science and Engineering, Beijing
Normal University, Beijing, 100875, P R China \\
 $^{4}$Testing and Analytical Center, Beijing
Normal University,
 Beijing, 100875, P R China\\
$^{5}$Department of Physics, \small{}Beijing Normal University,
Beijing, 100875, P R China \\
$^6$Center of Atomic and Molecular Nanosciences, Tsinghua
University, Beijing 100084, P R China }
\date{\today}
\begin{abstract}
  A multiple round quantum
dense coding (MRQDC) scheme based on the quantum phase estimation
algorithm is proposed.  Using an $m+1$ qubit system, Bob can
transmit $2^{m+1}$ messages to Alice, through manipulating only
one qubit and exchanging it between Alice and Bob for $m$ rounds.
The information capacity is enhanced to $m+1$ bits. We have
implemented the scheme in a three- qubit nuclear magnetic
resonance (NMR) quantum computer. The experimental results show a
good agreement between theory and experiment.
\end{abstract}
\maketitle \vspace{0.3cm}

\section{Introduction}\label{intro}
   Dense coding \cite{r1} can transmit more than one bit of
information by manipulating only one of two particles in an
entangled state. Initially, Alice has two particles in an
Bell-state \cite{s8}. She sends one of the two particles to Bob,
who manipulates the particle via one of four unitary operators,
namely $I$, $\sigma_{x}$, $i\sigma_{y}$ and $\sigma_{z}$, where
$\sigma_{x}$, $\sigma_{y}$, $\sigma_{z}$ and $I$ are the three
Pauli operators and the identity operator respectively, so as to
change the two-particle system into another of the four
Bell-states and then returns the processed particle to Alice.
 By determining the
Bell-state, Alice can read the encoded information. Compared with
the classical communication, in which one bit only transmits one
bit of information, dense coding doubles the information capacity.

  K. Mattle et al realized dense coding transmission with entangled
photon pairs \cite{r2}. They can transmit one of three messages by
manipulating only one of two entangled photons. X. Fang et al
demonstrated dense coding using a nuclear magnetic resonance
quantum computer \cite{r3}.  Four messages, i. e., two bits of
information, were transmitted in their scheme. X. Liu et al
proposed a general scheme for superdense coding among
multiparties, where more than two dimension systems were exploited
and the information capacity depended on the number of the
dimension of the individual system \cite{liu}. Grudka et al
proposed a symmetric multi-party superdense coding
scheme\cite{Grudka} in which the capacity increase is evenly
distributed among the participants.  K. Shimizu et al proposed a
dense coding scheme with enhanced information capacity through an
additional degree of freedom \cite{r15}. Dense coding for
continuous variables was also realized in experiments recently
\cite{xie03,xie02}. A. Harrow et al proposed superdense coding of
quantum states through generalizing dense coding to quantum states
\cite{Harrow}. I. P. Degiovanni et al exploited dense coding to
enhance the transmission capacity of quantum key distribution
(QKD). \cite{Degiovanni}

   Our scheme concentrates on dense coding using qubits.
   In previous scheme, the increase of
capacity is limited by the  dimension of the quantum system.
However, in quantum information processing, qubits are more
popularly used. Using an entangled qubit pair, the maximum
information dense coding can provide is 2 qubits. One common
feature of these dense coding schemes is that the qubits are only
transmitted one round. If we allow the qubit to travel between the
two users more than once, then it is possible to increase the
information capacity greatly. In this paper, we propose a multiple
round quantum dense coding (MRQDC) scheme, which can transmit
$2^{m+1}$ messages through manipulating only one qubit. The scheme
needs only an $m+1$ qubit system, in which only one qubit is
exchanged between Alice and Bob for $m$ rounds to encode
 $2^{m+1}$ messages. Alice decodes the $2^{m+1}$ messages
through the inverse of quantum Fourier transform (QFT). The MRQDC
scheme is the results of combining the quantum quantum phase
estimation  algorithm \cite{Nielsen} with the original dense
coding. In contrast, the usual dense coding scheme uses only a
single round, the maximum information it can transmit is only 2
bits.

  An feasible quantum system to demonstrate
the proposed quantum communication protocol is the nuclear
magnetic resonance (NMR) system,  and it has  become an important
arena for demonstrating quantum algorithms
\cite{Chuangprl,Vandersypen011,Weinstein,Long02,Peng,Lee}. Some
quantum communication protocols, such as the quantum
teleportation, quantum dense coding have  been demonstrated in NMR
quantum systems\cite{tele,dense,Fang}. We have  implemented the
proposed MRQDC scheme in a three qubit NMR quantum computer. The
effective-pure state is prepared using the spatial averaging
method.  Using this method, the scheme is implemented through only
one experiment, and the output is directly obtained by observing
the NMR spectrum.

\section{The Multiple Round Quantum Dense Coding Scheme}

\begin{figure}
\includegraphics[width=4in]{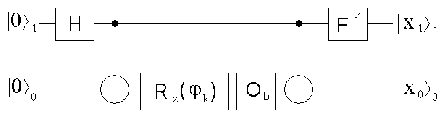}
\caption{The usual dense coding scheme, modified slightly for its
generalization. The horizontal lines denote qubits. $|0\rangle$
denotes the spin up state. $H$ denotes the Hadamard transform.
$R_{z}(\varphi_{k})$ denotes a phase manipulation, and
$R_{z}(\varphi_{k})=e^{i\varphi_{k}I_{z}}$, where
$\varphi_{k}=-k\pi$ $(k=0,1$). $O_{b}$ denotes a bit manipulation,
which can be chosen as $I$, or $\sigma_{x}$. $F^{-1}$ denotes the
inverse of quantum Fourier transform. $F^{-1}=H$ for the case of
one qubit. Time goes from left to right.
$|x_{1}\rangle_{1}|x_{0}\rangle_{0}$ is the output state, which
can be obtained through measuring qubits 1, and 0. }
\label{bennet}
\end{figure}

   The usual dense coding can be pictured as shown in Fig. \ref{bennet}.
   The horizontal lines denote qubits. $|0\rangle$
denotes the spin up state and  $H$ denotes the Hadamard transform.
Here the four different coding unitary operations are expressed as
$O_{b}R_{z}(\varphi_{k})$, where $R_{z}(\varphi_{k})$ denotes a
phase manipulation, and $O_{b}$ denotes a bit manipulation.
$O_{b}$ is chosen as $I$, or $\sigma_{x}$.
$R_{z}(\varphi_{k})=e^{i\varphi_{k}I_{z}}$, where
$\varphi_{k}=-k\pi$ $(k=0,1$).  $F^{-1}$ denotes the inverse of
quantum Fourier transform. For one qubit system, the Fourier
transform is just the Hadamard transformation, namely $F^{-1}=H$.
$|x_{1}\rangle_{1}|x_{0}\rangle_{0}$ is the output state. The four
transformation manipulations can be explicitly expressed as
$R_{z}(0)=I$, $R_{z}(-\pi)=-i\sigma_{z}$, $\sigma_{x}$, and
$R_{z}(-\pi)\sigma_{x}=\sigma_{y}$ respectively.   The four
manipulations correspond to the four output basis states
$|0\rangle_{1}|0\rangle_{0}$, $|1\rangle_{1}|0\rangle_{0}$,
$|0\rangle_{1}|1\rangle_{0}$, and $|1\rangle_{1}|1\rangle_{1}$,
respectively. Alice can determines the manipulation by making a
joint measurement on  qubits 1 and 0. During the dense coding
process, qubit 1 remains at Alice's site, and qubit 0 is sent to
Bob, then manipulated by Bob, and then sent back to Alice. The
Fourier transform is performed by Alice. Expressing the dense
coding lends clarity in generalizing it into a multiple round
dense coding scheme.

\begin{figure}
\includegraphics[width=6in]{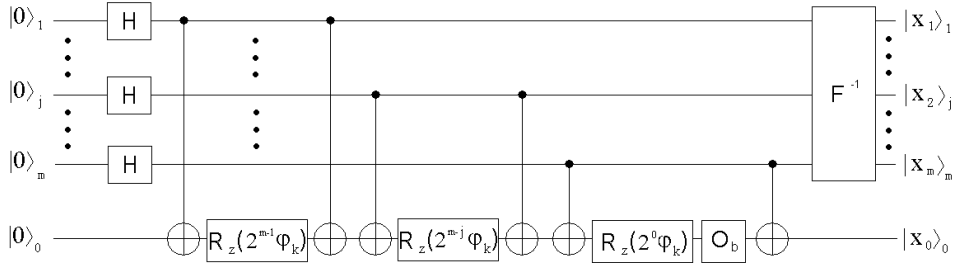}
\caption{The quantum network to implement the multiple round
quantum dense coding, obtained through generalizing the scheme
shown as Fig. \ref{bennet}.
$R_{z}(2^{m-j}\varphi_{k})=e^{i2^{m-j}\varphi_{k}I_{z}}$, where
$\varphi_{k}=-2k\pi/2^{m}$ $(k=0,1,2,\cdots, 2^{m}-1)$.
$|x_{1}\rangle_{1}\cdots|x_{j}\rangle_{j}\cdots|x_{m}\rangle_{m}|x_{0}\rangle_{0}$
is the output state, which can be obtained through measuring
qubits 1, 2, $\cdots$, $m$, and qubit 0.} \label{MRQDC}
\end{figure}

  Decomposing the encoding transformation into a phase rotation $R_{z}(\varphi_{k})$
  and a bit
  manipulation $O_{b}$ make it easier to generalize it into MRQDC.
  The phase manipulation can be viewed as quantum
manipulations, and the bit manipulations can be viewed as
classical manipulations,  just like decomposing information into a
quantum piece and a classical piece in Refs. \cite{lo1,lo2}. The
number of classical manipulation, either the identity $I$ or the
spin-flip $\sigma_x$ is fixed. The number of quantum
manipulations, however, can be extended to $2^{m}$, if the system
is extended to $m+1$ qubits. Combining with the two classical
manipulations, the MRQDC scheme can transmit $2^{m+1}$
manipulations,
    as  shown in Fig.
\ref{MRQDC}. The system has $m+1$ qubits, in which only qubit 0 is
 exchanged between Alice and Bob. For convenience,
qubit 0 is called the flying qubit, and the other $m$ qubits are
called the stationary qubits, which can be viewed as the quantum
memory. The flying qubit is exchanged between Alice and Bob to
encode the messages into the stationary qubit. The phase rotation
operations are
$R_{z}(2^{m-j}\varphi_{k})=e^{i2^{m-j}\varphi_{k}I_{z}}$ ($j=1$,
$2$, $\cdots$, $m$), where $\varphi_{k}=-2k\pi/2^{m}$ $(k=0$, $1$,
$2$, $\cdots$, $2^{m}-1)$. The output state can be rewritten as
$|k\rangle|x_{0}\rangle_{0}$. For example,
$|0\rangle|x_{0}\rangle_{0}=|0\rangle_{1}|0\rangle_{2}\cdots|0\rangle_{m}|x_{0}\rangle_{0}$,
$|1\rangle|x_{0}\rangle_{0}=|0\rangle_{1}|0\rangle_{2}\cdots|1\rangle_{m}|x_{0}\rangle_{0}$,
$\cdots$,
$|2^{m}-1\rangle|x_{0}\rangle_{0}=|1\rangle_{1}|1\rangle_{2}\cdots|1\rangle_{m}|x_{0}\rangle_{0}$.

   One of $2^{m+1}$ messages can be transmitted from Bob to
Alice through the following quantum communication scheme:

step 0:  Initially, Alice has $m+1$ qubits, each of which lies in
state $|0\rangle$.

step 1: Setting $j=1$. Alice starts MRQDC by manipulating qubit
$j$ and qubit 0.

step 2: Alice applies a Hadamard transform to qubit $j$, and then
a controlled-NOT gate (CNOT$_{j0}$) to qubits $j$ and 0 to
transform the two qubits into an entangled state. Then she sends
qubit 0 to Bob, who applies $R_{z}(2^{m-j}\varphi_{k})$ to qubit
0, noting
$R_{z}(2^{m-j}\varphi_{k})=R_{z}^{2^{m-j}}(\varphi_{k})$, i. e.,
repeating $R_{z}(\varphi_{k})$ for $2^{m-j}$ times. Then, he
returns qubit 0 to Alice. One should note that the two qubits are
still entangled. Alice applies $CNOT_{j0}$ to qubits $j$ and  0
for the second time to disentangle qubit 0 from qubit $j$.

step 3: Check if $j>m-1$. If it is not, put change $j$ to $j+1$
and go to step 2. If true, i. e., $j=m$, goto step 4.

step 4: Alice applies a Hadamard transform to qubit $m$, and then
CNOT$_{m0}$ to qubits $m$ and 0 to transform the two qubits into
an entangled state. Then she sends qubit 0 to Bob, who applies
$O_{b}R_{z}(\varphi_{k})$ to qubit 0. Then, he returns qubit 0 to
Alice. Alice applies $CNOT_{m0}$ to qubits $m$ and 0 for the
second time to disentangle qubit 0 from qubit $m$.

step 5: Making an inverse Fourier transform to the $m$ qubits, and
then make a joint measurement on qubits 0 to $m$. Using this
result, Alice reads out the information encoded by Bob.

Now let's analyze the working details of the MRQDC protocol. In
each round described in step 2, the effect on qubit $j (j<m)$ and
qubit 0 can be described as
\begin{equation}\label{1}
    T_{j}={\rm CNOT}_{j0}R_{z0}(2^{m-j}\varphi_{k}){\rm CNOT}_{j0}H_{j}.
\end{equation}
$T_{j}$ is the elementary operation for encoding in MRQDC. It can
be proved that
\begin{equation}\label{2}
    T_{j}|0\rangle_{j}|0\rangle_{0}=
    \frac{1}{\sqrt{2}}(|0\rangle_{j}+e^{-i2^{m-j}\varphi_{k}}|1\rangle_{j})|0\rangle_{0},
\end{equation}
where an irrelative overall phase factor has been ignored.

   Bob inputs the bit manipulation into the network in the last round described in step 4.
After the completion of $T_{m-1}T_{m-2}\cdots T_{1}$, the $m+1$
qubits are in a product state. Let
\begin{equation}\label{tm}
T_{m}={\rm CNOT}_{m0}O_{b}R_{z0}(\varphi_{k}){\rm CNOT}_{m0}H_{m},
\end{equation}
one obtains
\begin{equation}\label{rtm}
T_{m}|0\rangle_{m}|0\rangle_{0}=
\frac{1}{\sqrt{2}}(|0\rangle_{m}+e^{-i\varphi_{k}}|1\rangle_{m})O_{b}|0\rangle_{0},
\end{equation}
when $T_{m}$ is applied to qubit $m$ and qubit 0. Let
$T=T_{m}T_{m-1}\cdots T_{1}$,
\begin{equation}\label{3}
    T|0\rangle_{m}|0\rangle_{m-1}\cdots|0\rangle_{1}|0\rangle_{0}
    =\frac{1}{\sqrt{2^{m}}}\bigotimes_{j=1}^{m}
    (|0\rangle_{j}+e^{-i2^{m-j}\varphi_{k}}|1\rangle_{j})O_{b}|0\rangle_{0}
\end{equation}
is obtained. Eq. (\ref{3}) can be rewritten as
\begin{equation}\label{4}
T|0\rangle_{m}|0\rangle_{m-1}\cdots|0\rangle_{1}|0\rangle_{0}
    =\frac{1}{\sqrt{2^{m}}}\sum_{n=0}^{2^{m}-1}|n\rangle e^{i  2nk
    \pi/2^{m}}O_{b}|0\rangle_{0},
\end{equation}
using $\varphi_{k}=-2k\pi/2^{m}$. The decoding process is
implemented by the inverse of quantum Fourier transform, which
transforms Eq. (\ref{4}) into $|k\rangle O_{b}|0\rangle_{0}$
\cite{Nielsen}. By measuring the $m+1$ qubits, Alice obtains the
concrete value of $\varphi_{k}$, and know which manipulation Bob
has made. In this network, only the flying qubit is manipulated by
Bob, and $2^{m+1}$ messages can be transmitted from Bob to Alice.
When $m=1$, MRQDC returns to the original dense coding, where the
flying qubit is exchanged for only one round.

\section{Implementation of MRQDC in a 3-qubit NMR quantum system}
   Our experiments use a sample of Carbon-13 labelled
trichloroethylene (TCE) dissolved in d-chloroform. Data are taken
at room temperature with a Bruker DRX 500 MHz spectrometer.
$^{1}$H is the flying qubit, denoted by H0. The $^{13}$C directly
connecting to $^{1}$H is denoted as qubit 2, and the other
$^{13}$C is denoted as qubit 1. The two carbon nuclei are the
stationary qubits, denoted as C2 and C1, respectively. By setting
$\hbar=1$, the Hamiltonian of the three-qubit system is \cite{s10}
\begin{equation}\label{hamidun}
  H=-2\pi\nu_{1}I_{z}^{1}-2\pi\nu_{2}I_{z}^{2}-2\pi\nu_{0}I_{z}^{0}
  +2\pi J_{12}I_{z}^{1}I_{z}^{2}+2\pi J_{20 }I_{z}^{2}I_{z}^{0}
  +2\pi J_{10} I_{z}^{1}I_{z}^{0},
\end{equation}
where $I_{z}^{j}(j=0,1,2)$ are the matrices for the $z$-component
of the angular momentum of the spins. $\nu_{1}$, $\nu_{2}$,
$\nu_{0}$ are the resonance frequencies of C1, C2 and H0
respectively, and $\nu_{1}=\nu_{2}+904.4$Hz. The coupling
constants are measured to be $J_{12}=103.1$ Hz, $J_{20}=203.8$ Hz,
and $J_{10}=9.16$ Hz. The coupled-spin evolution between two spins
is denoted as
\begin{equation}\label{e2}
  [\tau]_{jl}=e^{-i2\pi J_{jl} \tau I_{z}^{j} I_{z}^{l}},
\end{equation}
where $l=0,1,2$, and $j\neq l$. $[\tau]_{jl}$ can be realized by
averaging the coupling constants other than $J_{jl}$ to
zero\cite{s15,Linden,Geen}.

  The initial effective-pure state $|000\rangle$ is prepared by
spatial averaging\cite{Cory,s9}.
   The following radio-frequency (rf) pulse and gradient pulse
sequence
\begin{eqnarray}
&&[\pi/4]_{x}^{1,2}-[1/2J_{12}]_{12}-[-5\pi/6]_{y}^{1,2}-[\alpha]_{x}^{0}-[{\rm
grad}]_{z}
-[\pi/4]_{y}^{0}-[9/2J_{23}]_{23}-\nonumber\\
&&[1/2J_{13}]_{13}-[\pi/4]_{y}^{0}-[{\rm grad}]_{z}
-[\pi/4]_{y}^{0}-[9/4J_{23}]_{23}-[1/4J_{13}]_{13}-[\pi/4]_{x}^{0}-[{\rm
grad}]_{z},
\end{eqnarray}
transforms the system from the equilibrium state
\begin{equation}\label{equ}
  \rho_{eq}=\gamma_{C}(I_{z}^{1}+ I_{z}^{2})+\gamma_{H}I_{z}^{0},
\end{equation}
to
\begin{equation}\label{2}
  \rho_{0}=I_{z}^{1}/2+I_{z}^{2}/2+I_{z}^{0}/2+I_{z}^{1}I_{z}^{2}+I_{z}^{2}I_{z}^{0}+
  I_{z}^{1}I_{z}^{0}+2I_{z}^{1}I_{z}^{2}I_{z}^{0},
\end{equation}
where an overall phase factor has been ignored.
$[\pi/4]_{x}^{1,2}$ denotes the $\pi/4$ pulse exciting C1 and C2
simultaneously along x-axis. $[\pi/4]_{y}^{0}$ denotes the $\pi/4$
 spin-selective pulse for $^{1}$H along y-axis. $\gamma_{C}$ and
$\gamma_{H}$ denotes the gyromagnetic ratio of $^{13}$C and
$^{1}$H. $\alpha=\arccos(-\gamma_{C}\sqrt{6}/\gamma_{H})$.
$\rho_{0}$ is equivalent to $|000\rangle$ in NMR experiments
\cite{Knill1,Somarooprl991,Zhang10}. We find that the compound
operations
\begin{eqnarray}
 {\rm CNOT}_{10}R_{z0}(2\varphi_{k}){\rm CNOT}_{10}
&=&[\frac{-2\varphi_{k}}{\pi J_{10}}]_{10},\label{C13} \\
 {\rm CNOT}_{20}R_{z0}(\varphi_{k}){\rm CNOT}_{20}
&=&[\frac{-\varphi_{k}}{\pi J_{20}}]_{20},\label{C23}
\end{eqnarray}
make the network easier to realize. The Hadamard transform
simultaneously applied to C1 and C2, denoted by $H^{1,2}$
respectively, is realized by pulse sequence
$[-\pi/2]_{y}^{1,2}-[\pi]_{x}^{1,2}$. The inverse of QFT
$F^{-1}=H^{1}I_{|11\rangle}^{-\pi/2}H^{2}SWAP$. $H^{2}$ denotes
the Hadamard transform applied to C2, and is realized by
$[\pi/4]_{y}^{1,2,0}-[\pi]_{z}^{2}-[-\pi/4]_{y}^{1,2,0}$, where
$[\pi]_{z}^{2}$ denotes the  $z$-pulse, which can be realized by
rf pulses \cite{Linden}. $H^{1}=H^{1,2}H^{2}$, using
$H^{2}H^{2}=I$. $I_{|11\rangle}^{-\pi/2}$ denotes the controlled-
phase shift gate, and is realized by
$[1/4J_{12}]-[-\pi/2]^{1,2}_{y}-[\pi/4]^{1,2}_{x}-[\pi/2]^{1,2}_{y}
 $\cite{Weinstein,Longpla,zhangcp}. The SWAP gate can be counteracted by another one
through optimizing the network \cite{Nielsen}.

\begin{figure}
\includegraphics[width=6in]{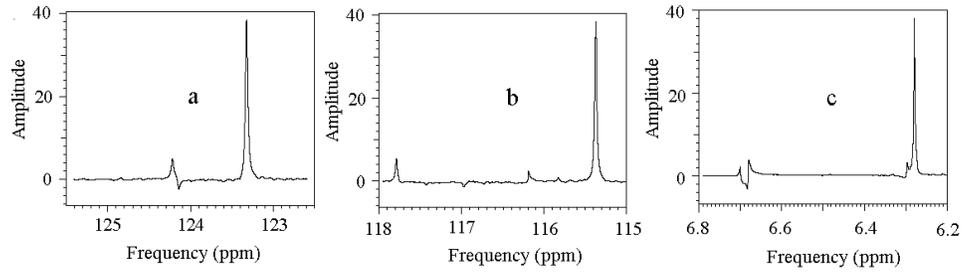}
\caption{The NMR spectra of C1, C2, and H0, shown as Figs. (a-c),
when the three-spin system lies in effective- pure state
$|000\rangle$ via readout pulses selective
 for C1, C2, and H0, respectively. The amplitude
has arbitrary units. When the system lies in $|000\rangle$, only
one NMR peak appears in each spectrum if a spin selective readout
pulse is applied.} \label{ref}
\end{figure}

    Our experiments results are represented as NMR spectra obtained by
 spin-selective readout pulses $[\pi/2]_{y}^{1}$,
 $[\pi/2]_{y}^{2}$ and $[\pi/2]_{y}^{0}$, for the three spins, respectively.
Figs. \ref{ref} (a-c) show the NMR spectra of C1, C2, and H0, when
the three-spin system lies in effective- pure state $|000\rangle$,
via the readout pulses selective for C1, C2, and H0, respectively.
It is seen that only one NMR peak appears in each spectrum. The
last gradient pulse in preparing the effective- pure state has
killed the non-diagonal elements of the density matrix, the three
spectra in Fig. \ref{ref} are sufficient to reconstruct the
density matrix of $|000\rangle$ using the state tomography
technique. The signals in Figs. \ref{ref}(a-b) are chosen as the
reference signals to calibrate the phase of the signals in the
following carbon spectra in order to make the phases of NMR
signals meaningful \cite{Jones}.

\begin{figure}
\includegraphics[width=6in]{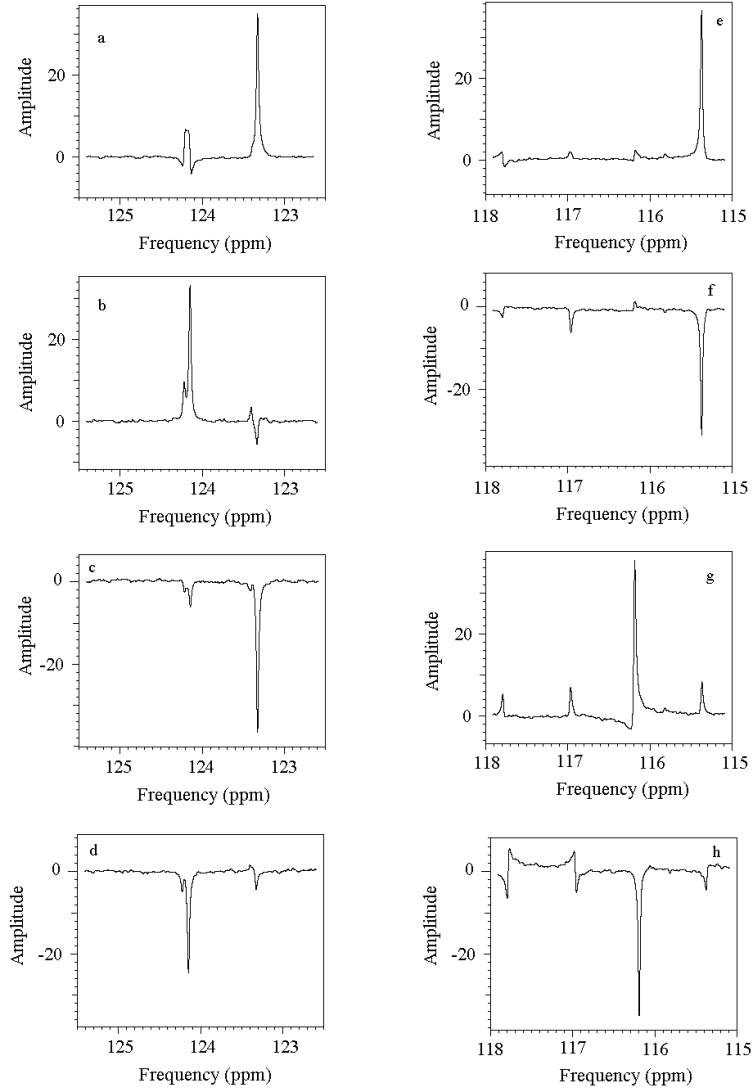}
\caption{The spectra of C1 (shown by the left column) and C2
(shown by the right column) obtained through $[\pi/2]_{y}^{1}$ for
C1 and $[\pi/2]_{y}^{2}$ for C2  after the completion of multiple
round quantum dense coding (MRQDC). Figs. (a-d) and Figs. (e-h)
correspond to states $|00\rangle_{1,2}|0\rangle_{0}$,
$|01\rangle_{1,2}|0\rangle_{0}$, $|10\rangle_{1,2}|0\rangle_{0}$,
and $|11\rangle_{1,2}|0\rangle_{0}$, respectively.}
\label{results}
\end{figure}

   MRQDC starts with $|000\rangle$. Without loss of generality and considering the
convenience in experiments, we demonstrate the case of $O_{b}=I$.
The network shown as Fig. \ref{MRQDC} transforms
$|00\rangle_{1,2}|0\rangle_{0}$ to
$|00\rangle_{1,2}|0\rangle_{0}$, $|01\rangle_{1,2}|0\rangle_{0}$,
$|10\rangle_{1,2}|0\rangle_{0}$, and
$|11\rangle_{1,2}|0\rangle_{0}$, respectively corresponding to the
four different manipulations $R_{z}(0)=I$ (identity manipulation
or no manipulation), $R_{z}(-\pi/2)$, $R_{z}(-\pi)$, and
$R_{z}(-3\pi/2)$. By applying the spin-selective readout pulses,
we obtain the spectra of C1 shown in Figs.\ref{results}(a-d), and
spectra of C2 shown as Figs.\ref{results}(e-h), corresponding to
the above four manipulations, respectively\cite{s9}. In
experiments, we have noted that if no readout pulse is applied,
the amplitudes of peaks in each spectrum is so small that they can
be ignored. Due to imperfection of the pulse sequence in NMR, the
inhomogeneity in the magnetic field and the decoherence time
limit, there are errors in the experiment. The overall
experimental errors are less than 20\% (barring Fig. \ref{results}
(b) and (h)). The accuracies are acceptable in NMR
\cite{Weinstein,Lee}.

   The case of $O_{b}=\sigma_{x}$ can be implemented in a similar
way, where $\sigma_{x}$ can be realized by pulse $[\pi]_{x}^{0}$.
The output states are $|00\rangle_{1,2}|1\rangle_{0}$,
$|01\rangle_{1,2}|1\rangle_{0}$, $|10\rangle_{1,2}|1\rangle_{0}$,
and $|11\rangle_{1,2}|1\rangle_{0}$, corresponding to the four
different manipulations $\sigma_{x}R_{z}(0)=\sigma_{x}$,
$\sigma_{x}R_{z}(-\pi/2)$, $\sigma_{x}R_{z}(-\pi)$, and
$\sigma_{x}R_{z}(-3\pi/2)$, respectively.

\section{Conclusion}
   The multiple round quantum dense coding has been proposed. It enables two parties to transmit $m+1$
   qubits of information by exchanging only a single qubit in multiple rounds. We have also
implemented the protocol in a three- qubit NMR quantum computer.
The scheme combines the quantum  phase estimation algorithm with
the original dense coding protocol. The phase estimation algorithm
is a very powerful tool in quantum computing and quantum
information processing. It has been successfully used in clock
synchronization that outperforms classical counterpart
exponentially \cite{Chuang}. By applying the phase estimation
algorithm to the dense coding, the information capacity can be
enhanced to arbitrary amount of bits in a systematic manner. The
elementary operations for encoding can be realized directly
through the couplings between the flying qubits and the stationary
qubit. Consequently, the scheme is easy to carry out in practice.
In practical applications, photons are the natural first choice as
the flying qubits. The entangled atom- photon system prepared by
Blinov et al \cite{Blinov} is a potential system  for the
realization of MRQDC, where the photon acts as the flying qubit,
because it is easy to be transmitted between Alice and Bob. The
$m$ atom qubits act as the stationary qubits. $m+1$ bits of
information can be transmitted through manipulating the photon by
Bob, and exchanging it between Alice and Bob. Moreover, one can
find that different photons can be used in the different rounds.
This fact will greatly simplify the process of the implementation
of the scheme.

  The quantum advantage of MRQDC mainly lies in the fact that
MRQDC transmits messages using quantum manipulations. The
information is encoded in the phase of a quantum state described
as Eq. (2). The measurement for the flying qubit during the
process of transmission does not obtain any information of the
message, so that the communication is secure. This propery of
dense coding has been applied to quantum key distribution in
several quantum key distribution protocols  \cite{rmp,deng}.

\section{Acknowledgement}
  This work is supported by the National Natural Science
Foundation of China under Grant No. 10374010, 60073009, 10325521,
the National Fundamental Research Program Grant No. 001CB309308,
the Hang-Tian Science Fund, the SRFDP program of Education
Ministry of China, and China Postdoctoral Science Foundation. We
are grateful for Dr. Fuguo Deng for the fruitful discussions.

\newpage


\end{document}